\documentclass[twocolumn, superscriptaddress,pre,notitlepage]{revtex4-1}
\usepackage[T1]{fontenc}
\usepackage{amsmath}
\usepackage{amssymb}
\usepackage{graphicx}

\makeatletter
\usepackage{bm}\usepackage{amsfonts}
\usepackage{color}

\makeatother

\begin{document}
\title{A Shortcut to Finite-time Memory Erasure}
\author{Geng Li}
\affiliation{Graduate School of China Academy of Engineering Physics, Beijing 100193,
China}
\affiliation{School of Systems Science, Beijing Normal University, Beijing 100875,
China}
\author{Hui Dong}
\email{hdong@gscaep.ac.cn}

\affiliation{Graduate School of China Academy of Engineering Physics, Beijing 100193,
China}
\begin{abstract}
To achieve fast computation, it is crucial to reset the memory to
a desired state within a limited time. However, the inherent delay
in the system's response often prevents reaching the desired state
once the control process is completed in finite time. To address this
challenge, we propose a shortcut strategy that incorporates an auxiliary
control to guide the system towards an equilibrium state that corresponds
to the intended control, thus enabling accurate memory reset. Through
the application of thermodynamic geometry, we derive an optimal shortcut
protocol for erasure processes that minimizes the energy cost. This
research provides an effective design principle for realizing the
finite-time erasure process while simultaneously reducing the energy
cost, thereby alleviating the burden of heat dissipation.
\end{abstract}
\maketitle
\emph{Introduction.} -- Memory erasure is an essential step in computations
with an unavoidable energy cost. Landauer's principle posts a fundamental
lower bound on the energy cost of erasing a one-bit memory carried
out infinitely slowly \citep{Landauer1961,Bennett1973,Parrondo2015}.
However accelerating computing processes typically requires to complete
memory erasure in finite time. There is an inevitable trade-off between
erasure speed and accuracy when rapidly initializing the memory system.
In such a finite-time thermodynamic process, a system often does not
immediately respond to external perturbation, resulting in a lag between
the final state and the desired equilibrium state \citep{Pearlman1989,Wood1991}.
This systematic state lag poses a significant challenge in initializing
the system to the target state promptly \citep{Dillenschneider2009,Berut2012,Jun2014,Berut2015,Proesmans2020,Proesmans2020a,Boyd2022},
and also leads to a substantial increase in energy costs when the
erasure time is reduced \citep{Kawai2007,GomezMarin2008a,Vaikuntanathan2009,Horowitz2009,Diana2013,Zulkowski2014,Zulkowski2015,Dago2021,Vu2022,Ma2022,Rolandi2022,Rolandi2023}.
The quest to achieve rapid erasure with minimal energy costs drives
us to design alternative memory erasure strategies.

\begin{figure}[!htp]
\includegraphics[width=8.5cm]{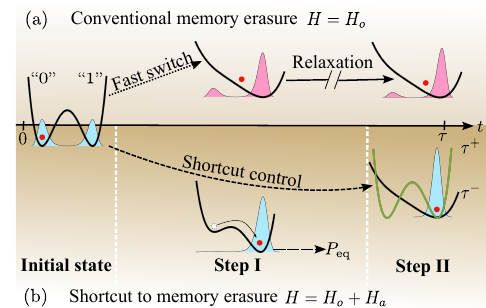} \caption{Schematic of finite-time memory erasure. The one-bit memory is modeled
by a physical double-well system. In the initial state, the system
stores the state's information $``0"$ or $``1"$. The task of memory
erasure is to reset the system into the blank state $``1"$. (a) Conventional
memory erasure scheme with the Hamiltonian $H_{o}$. The double-well
potential is fast switched to fulfill classical memory reset in finite
time $\tau$. A systematic state lag is accumulated between the system's
current state and its corresponding equilibrium state. Therefore,
a slow relaxation process is needed to reach the desired erasure accuracy.
(b) Shortcut to memory erasure with the Hamiltonian $H=H_{o}+H_{a}$.
In the step I, an auxiliary Hamiltonian $H_{a}$ is added to escort
the system evolving along the instantaneous equilibrium path of the
original Hamiltonian $H_{o}$ within finite time. The system is finally
expelled into the blank state $``1"$ without accumulation of the
state lag. In the step II, the potential is quenched to the double-well
form for later computations.}
\label{Schematic}
\end{figure}

Much effort has been devoted to reduce the state lag developed in
nonequilibrium driving processes \citep{Miller2000,Jarzynski2002,Vaikuntanathan2008,Minh2009,Martinez2016,Patra2017,Li2017}.
As a promising candidate, shortcut to isothermality was proposed as
a finite-time driving strategy to escort the system evolving along
a series of instantaneous equilibrium states \citep{Li2017} with
the application to realize fast transitions between equilibrium states
\citep{Albay2019,Albay2020}, improve free energy calculations \citep{Li2019,Li2021},
and design Brownian heat engines \citep{Martinez2017,Plata2020,Chen2022,Zhao2022}.
As shown in Fig.$\ $\ref{Schematic}, we adopt such shortcut strategy
to design a finite-time erasure protocol to reset a classical memory.
And the thermodynamic geometric approach \citep{Li2022,Li2023,Chittari2023}
is used to design the optimal protocol with the minimal energy cost.

\emph{Shortcuts to memory erasure.} -- The memory as a simple binary
system can be simplified as a particle in a bistable potential well.
Here, we consider a one-bit memory system modeled by a Brownian particle
in a double-well potential $U_{o}(x,\vec{\lambda})=kx^{4}-A\lambda_{1}x^{2}-B\lambda_{2}x$,
where $x$ represents the coordinate. And $k$, $A$, and $B$ are
constant coefficients introduced to define the dimensionless variables
$\vec{\lambda}(t)\equiv(\lambda_{1},\lambda_{2})$ as time-dependent
control parameters. The evolution of the system is described by Hamiltonian
$H_{o}(x,p,\vec{\lambda})\equiv p^{2}/(2m)+U_{o}(x,\vec{\lambda})$
with $p$ as the momentum and $m$ as the mass of the particle. The
memory is encoded by mapping the microstate $x$ into two macrostates.
If the particle is in the left well $(x<0)$, the system is in macrostate
``$0$''. Conversely, if the particle is in the right well $(x>0)$,
the system is in macrostate ``$1$'', which serves as the blank state
for storing memory.

The system is in contact with a thermal reservoir with a constant
temperature $T$. The barrier separating the double well is assumed
to be much larger than the thermal fluctuation so that the memory
can be considered stable \citep{Dillenschneider2009,Jun2014}.

The particle is initially in an equilibrium microstate with the control
parameters $\vec{\lambda}(0)=(1,0)$ and has the same probability
$1/2$ to stay in each macrostate ($0$ or $1$). The entropy is $S=k_{B}\ln2$
for the initial microstate of the particle with $k_{B}$ being the
Boltzmann constant. During the Landauer's quasi-static erasure process
\citep{Landauer1961,Bennett1973,Parrondo2015}, the particle is expelled
into macrostate ``$1$'' by changing control parameters $\vec{\lambda}$
slowly to maintain the equilibrium state $P_{\mathrm{eq}}=\exp[\beta(F-H_{o})]$,
where $F\equiv-\beta^{-1}\ln[\iint dxdp\exp(-\beta H_{o})]$ is the
free energy with the inverse temperature $\beta\equiv1/(k_{B}T)$.
And the entropy for the final state is $S=0.$ Such reduction of the
entropy makes the memory erasure a logically irreversible process.
Once the control parameters $\vec{\lambda}$ are tuned with finite
rate as illustrated in Fig.$\ $\ref{Schematic} (a), the system is
driven to a nonequilibrium state typically with a lag, resulting in
a residual entropy $S>0$. And additional relaxation is required to
achieve the desired accuracy with the sacrifice of longer time \citep{Maruyama2009,Parrondo2015,Bechhoefer2020}.

To meet the need of fast erasure and reduce the lag, we adopt the
shortcut scheme where an auxiliary Hamiltonian $H_{a}(x,p,t)$ is
supplemented to escort the system evolving along the instantaneous
equilibrium state $P_{\mathrm{eq}}$ during the finite-time erasure
process with boundary conditions $H_{a}(0)=H_{a}(\tau)=0$. The probability
distribution $P(x,p,t)$ of the microstate follows the Kramers equation,

\begin{equation}
\frac{\partial P}{\partial t}=-\frac{\partial}{\partial x}(\frac{\partial H}{\partial p}P)+\frac{\partial}{\partial p}(\frac{\partial H}{\partial x}P+\gamma\frac{\partial H}{\partial p}P)+\frac{\gamma}{\beta}\frac{\partial^{2}P}{\partial p^{2}}],\label{eq:kramerseq}
\end{equation}
where $H\equiv H_{o}+H_{a}$ is the total Hamiltonian and $\gamma$
is the dissipation coefficient. Our operation of memory erasure, illustrated
in Fig. 1 (b), consists two steps as follows.

\textbf{Step I, Shortcut Erasure. }The particle is expelled to the
right well $(x>0)$ in the finite-time interval $t\in[0,\tau]$, by
raising the left well, illustrated as the step I in Fig.$\ $\ref{Schematic}
(b). The control parameters are tuned from $\vec{\lambda}(0)$ to
$\vec{\lambda}(\tau)=(0,1)$. With the strategy of shortcuts to isothermality
\citep{Li2017}, the system evolves along the path of instantaneous
equilibrium states, and reaches the final equilibrium state $P=P_{\mathrm{eq}}(\tau)$
with the control parameters $\vec{\lambda}(\tau)$. The requirement
for the auxiliary Hamiltonian $H_{a}$ follows as

\begin{equation}
\frac{\gamma}{\beta}\frac{\partial^{2}H_{a}}{\partial p^{2}}-\text{\ensuremath{\frac{\gamma p}{m}}}\frac{\partial H_{a}}{\partial p}+\frac{\partial H_{a}}{\partial p}\frac{\partial H_{o}}{\partial x}-\frac{p}{m}\frac{\partial H_{a}}{\partial x}=\frac{dF}{dt}-\frac{\partial H_{o}}{\partial t}.\label{eq:additionalHamiltonianeq}
\end{equation}
The auxiliary Hamiltonian is proved to have the form $H_{a}(x,p,t)=\dot{\vec{\lambda}}\cdot\vec{f}(x,p,\vec{\lambda})$
with the boundary condition $\dot{\vec{\lambda}}(0)=\dot{\vec{\lambda}}(\tau)=0$.

\textbf{Step II, Potential Quench. }For later computation purpose,
the potential is reset to the double-well form. Such operation is
realized by quenching the control parameters from $\vec{\lambda}(\tau^{-})=(0,1)$
to the initial value $\vec{\lambda}(\tau^{+})=(1,0)$ at the time
$t=\tau$ with an instantaneous change of the system Hamiltonian $H_{o}(\tau^{-})\to H_{o}(\tau^{+})$.
Here, $\tau^{-}$ and $\tau^{+}$ denote the time before and after
the potential quench respectively.

After these two steps, the memory is erased to the blank state and
the system is reset within finite time $\tau$ to allow the later
usage.

\emph{Geometric erasure protocol.} -- Energy cost is inevitable in
a finite-time erasure process. We employ the geometric approach \citep{Li2022,Li2023}
to derive the optimal erasure protocol with minimal energy costs.
In the shortcut scheme with the total Hamiltonian $H=H_{o}+H_{a}$,
the work performed in the erasure process with duration $\tau$ is
$W\equiv\langle\int_{0}^{\tau}\partial H/\partial tdt\rangle_{\mathrm{eq}},$
which is explicitly obtained as

\begin{equation}
W_{s}=\Delta F+\gamma\int_{0}^{\tau}dt\langle(\frac{\partial H_{\mathrm{a}}}{\partial p})^{2}\rangle_{\mathrm{eq}},\label{eq:approxtwork}
\end{equation}
where $\langle\cdot\rangle_{\mathrm{eq}}\equiv\iint dxdp[\cdot]P_{\mathrm{eq}}$.
Since the potential quench in the Step II is realized instantaneously,
the system distribution remains unchanged. And the work done in this
process follows as $W_{q}=\iint dxdp(H_{o}(\tau^{+})-H_{o}(\tau^{-}))P_{\mathrm{eq}}(\tau)$.

With the explicit form of the auxiliary Hamiltonian $H_{a}=\dot{\vec{\lambda}}\cdot\vec{f}$,
the irreversible energy cost of the erasure process is written as

\begin{eqnarray}
W_{\mathrm{irr}} & \equiv & W-\Delta F-W_{q}\nonumber \\
 & = & \gamma\sum_{\mu\nu}\int_{0}^{\tau}dt\dot{\lambda}_{\mu}\dot{\lambda}_{\nu}\langle\frac{\partial f_{\mu}}{\partial p}\frac{\partial f_{\nu}}{\partial p}\rangle_{\mathrm{eq}},\label{eq:appirrwork}
\end{eqnarray}
with the total work $W=W_{s}+W_{q}.$ Here $W_{q}$ is excluded from
the irreversible energy cost $W_{\mathrm{irr}}$ in Eq.$\ $(\ref{eq:appirrwork})
since the potential quench in the Step II is a reversible process.
In the space of the control parameters $\vec{\lambda}$, a semi-positive
metric can be defined as $g_{\mu\nu}\equiv\gamma\langle\partial f_{\mu}/\partial p\partial f_{\nu}/\partial p\rangle_{\mathrm{eq}}$
on a Riemannian manifold. The shortest curve connecting two given
endpoints in this parametric space is the geodesic line with the distance
described by the thermodynamic length \citep{Salamon1983,Crooks2007,Sivak2012,Chen2021}
$\mathcal{L}\equiv\int_{0}^{\tau}dt\sqrt{\sum_{\mu\nu}\dot{\lambda}_{\mu}\dot{\lambda}_{\nu}g_{\mu\nu}}$,
which provides a lower bound for the irreversible energy cost $W_{\mathrm{irr}}\ge\mathcal{L}^{2}/\tau.$
Therefore, the geodesic line serves as the optimal erasure protocol
with minimal energy costs.

In the shortcut scheme, the auxiliary Hamiltonian $H_{a}$ can be
solved from Eq.$\ $(\ref{eq:additionalHamiltonianeq}) with the given
original Hamiltonian $H_{o}$. However, the form of the auxiliary
Hamiltonian $H_{a}$ generally depends on the particle's momentum
\citep{Li2017,Li2021,Li2023}. The demand of constantly monitoring
the particle's velocity makes it difficult for implementation of momentum-dependent
terms in experiment \citep{GueryOdelin2019,GueryOdelin2023}. Here
we design a variational auxiliary control $H_{a}^{*}=\dot{\vec{\lambda}}\cdot\vec{f}^{*}(x,p,\vec{\lambda})$
that is used to replace the exact auxiliary Hamiltonian $H_{a}$,
where $\vec{f}^{*}$ represents an approximation to the function $\vec{f}$
to ensure the minimization of a variational functional as

\begin{eqnarray}
\mathcal{G}(H_{a}^{*}) & = & \int dxdp(\frac{\gamma}{\beta}\frac{\partial^{2}H_{a}^{*}}{\partial p^{2}}-\frac{\gamma p}{m}\frac{\partial H_{a}^{*}}{\partial p}+\frac{\partial H_{o}}{\partial x}\frac{\partial H_{a}^{*}}{\partial p}\nonumber \\
 &  & -\frac{p}{m}\frac{\partial H_{a}^{*}}{\partial x}+\frac{\partial H_{o}}{\partial t}-\frac{\mathrm{d}F}{\mathrm{d}t})^{2}\mathrm{e}^{-\beta H_{o}}.\label{eq:noncons}
\end{eqnarray}
Such variational functional is defined to ensure the minimum discrepancy
between evolutions governed by the approximate Hamitonian $H_{a}^{*}$
and the exact Hamitonian $H_{a}$. And the variational auxiliary Hamiltonian
for the Brownian particle takes the form $H_{\mathrm{a}}^{*}=\text{\ensuremath{\sum_{\mu=1}^{2}\dot{\lambda}_{\mu}f_{\mu}^{*}(x,p,\vec{\lambda})}}$
with $f_{1}^{*}=a_{4}xp+a_{3}p+a_{2}x^{2}+a_{1}x,$ and $f_{2}^{*}=b_{4}xp+b_{3}p+b_{2}x^{2}+b_{1}x$.
Here $a_{n}\equiv a_{n}(\vec{\lambda})$ and $b_{n}\equiv b_{n}(\vec{\lambda})$
with $n=1,2,3,4$ are functions to be determined through the variational
procedure. With the adoption of the shortcut scheme under the total
Hamiltonian $H=H_{o}+H_{a}^{*}$, the system can be escorted along
a series of near-equilibrium state $P_{\mathrm{eq}}^{*}\approx P_{\mathrm{eq}}$.

With the operation of a gauge transformation $X=x$ and $P=p+m\partial H_{a}^{*}/\partial p$,
we obtain an equivalent process

\begin{equation}
\dot{X}=\frac{P}{m},\ \dot{P}=-\frac{\partial U_{o}}{\partial X}-\frac{\partial U_{a}}{\partial X}-\gamma\dot{X}+\xi(t),\label{eq:equivproc}
\end{equation}
where $\xi(t)$ is the Gaussian white noise and the auxiliary potential
$U_{a}(X,t)=C_{2}(t)X^{2}+C_{1}(t)X$ is momentum-independent. Detailed
derivations of the equivalent process in Eq.$\ $(\ref{eq:equivproc})
and the lengthy expressions of $C_{1}(t)$ and $C_{2}(t)$ are presented
in Supplemental Material \citep{Supple}. In this equivalent process,
the system's distribution follows a fixed pattern 
\begin{equation}
P_{f}(X,P,t)=\exp\left\{ \beta[F-\frac{1}{2m}(P-m\partial H_{a}^{*}/\partial P)^{2}-U_{o}]\right\} .\label{eq:fixedpattern}
\end{equation}
With the boundary conditions $\dot{\vec{\lambda}}(0)=\dot{\vec{\lambda}}(\tau)=0$,
the extra term $m\partial H_{a}^{*}/\partial P$ in Eq.$\ $(\ref{eq:fixedpattern})
vanishes and the system's distribution $P_{f}$ returns to the instantaneous
equilibrium distribution $P_{\mathrm{eq}}^{*}$ at the beginning $t=0$
and end $t=\tau$ of the driving process.

\begin{figure}[!htp]
\includegraphics[width=8.5cm]{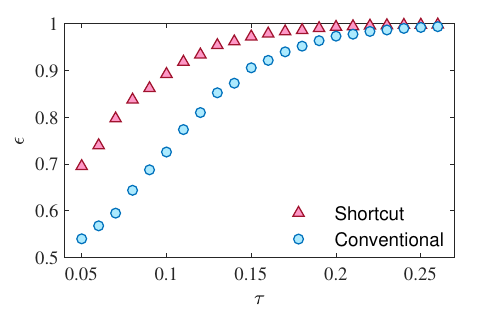} \caption{The erasure accuracy $\epsilon$ of the shortcut scheme (red triangles)
and the conventional erasure scheme (blue circles). The erasure accuracy
is defined as the probability of particles ending in the right well
$\epsilon\equiv\int_{0}^{\infty}P(x,\tau)dx$. During the erasure
process, the shortcut scheme adopts the Hamiltonian $H=H_{o}+U_{a}$
while the conventional scheme takes the Hamiltonian $H'=H_{o}$. In
the simulation, we choose the parameters as $k=4$, $A=8$, $B=16$,
$k_{B}T=1$, $\gamma=1$, and $m=0.01$ and the straightforward control
protocol as $\lambda_{1}^{s}(t)=0.5+0.5\cos(\pi t/\tau)$ and $\lambda_{2}^{s}(t)=0.5-0.5\cos(\pi t/\tau)$.
The shortcut scheme always ensures high erasure accuracy while the
conventional erasure scheme only achieves considerable erasure accuracy
at long erasure duration.}
\label{Accuracy}
\end{figure}

\emph{Reset errors.} -- The conventional approach of memory erasure
is to tune the system Hamiltonian $H_{o}$ via control parameters
$\vec{\lambda}$. The memory state evolves accordingly yet with a
lag, which induces reset errors especially for the case with short
erasure time $\tau$. The shortcut scheme with the Hamiltonian $H=H_{o}+U_{a}$
effectively fulfills the demands for both erasure speed and accuracy,
without introducing any additional freedom of control.

We numerically obtain the shortcut scheme with the Hamiltonian $H=H_{o}+U_{a}$
and compare it with the conventional scheme with the Hamiltonian $H'=H_{o}$
in a one-bit memory erasure process. In simulations, we choose the
parameters as $k=4$, $A=8$, $B=16$, $k_{B}T=1$, $\gamma=1$, and
$m=0.01$. See Supplemental Material for more details about the simulation
\citep{Supple}. A straightforward control protocol is selected as
$\lambda_{1}^{s}(t)=0.5+0.5\cos(\pi t/\tau)$ and $\lambda_{2}^{s}(t)=0.5-0.5\cos(\pi t/\tau)$
to continuously meet the boundary conditions. The erasure accuracy
is defined as the relative number of particles reaching the right
well at the end of the erasure process $\epsilon\equiv\int_{0}^{\infty}P(x,\tau)dx$.
Figure \ref{Accuracy} shows the erasure accuracy $\epsilon$ for
different erasure durations $\tau$. The accuracy of the shortcut
scheme is superior to that of the conventional scheme over different
durations. Especially in short erasure time, the conventional scheme
almost fails while the accuracy of the shortcut scheme still maintains
high level.

\begin{figure}[!htp]
\includegraphics[width=8.5cm]{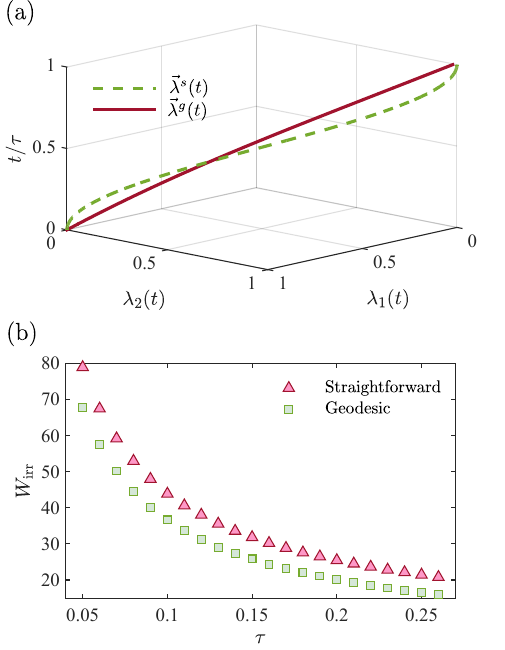} \caption{(a) Geodesic protocols for the memory erasure process with the shortcut
scheme. The control parameters vary form the initial value $\vec{\lambda}(0)=(1,0)$
to the final value $\vec{\lambda}(\tau)=(0,1)$. The dashed lines
represent the straightforward protocol while the solid lines represent
the geodesic protocol. The geodesic protocol is obtained by solving
the geodesic equation corresponding to the metric $g_{\mu\nu}$. (b)
The irreversible energy cost of the straightforward protocol (red
triangles) and the geodesic protocol (green squares). The energy cost
of erasure processes given by the geodesic protocol is lower than
that from the straightforward protocol.}
\label{Energycost}
\end{figure}

\emph{Energy cost optimization.} -- In the shortcut scheme, the auxiliary
Hamiltonian $H_{a}^{*}$ is obtained once the control protocol $\vec{\lambda}(t)$
with $t\in[0,\tau]$ is specified for the conventional erasure process.
Among these protocols, one with the minimum energy cost can be found
with our geometric methods. We test the geometric approach for minimizing
the energy cost of the erasure process. The geodesic protocol is obtained
by numerically solving the geodesic equation in the parametric space
$\vec{\lambda}$ with the metric $g_{\mu\nu}$. The details on the
form of the metric and the geodesic equation are presented in the
Supplementary Materials \citep{Supple}. Figure \ref{Energycost}
(a) shows the difference between the geodesic protocol and the above
straightforward protocol $\vec{\lambda}^{s}(t)$. In Fig. \ref{Energycost}
(b), we compare the irreversible energy cost of the geodesic protocol
(green squares) with that of the straightforward protocol (red triangles).
The irreversible energy cost $W_{\mathrm{irr}}$ given by the geodesic
protocol is lower than that from the straightforward protocol. Such
observation proves the geometric approach for the shortcut scheme
to be effective. With the application of tools from Riemannian geometry
\citep{Berger2007}, the procedure of searching the optimal erasure
protocol is largely simplified with our two-step processes.

\emph{Conclusions.} -- In summary, we have developed a shortcut strategy
to realize a finite-time one-bit memory erasure with minimal energy
costs. In the shortcut scheme, an auxiliary control have been introduced
to steer the system evolving along the path of instantaneous equilibrium
states. We have employed the variational procedure to remove the momentum-dependent
terms and derive an accessible auxiliary control. The irreversible
energy cost of the erasure process have been minimized by adopting
the geometric method that connects the optimal erasure protocol with
the geodesic line in a Riemannian manifold. This property helps us
to solve the optimal protocol by using methods developed in geometry.
Numerical results have verified that the shortcut strategy can largely
improve the accuracy of finite-time memory erasure without additional
control means. Our strategy shall provide an effective design principle
for finite-time memory erasure with low energy costs.

Recently, much effort has been devoted to realize the one-bit memory
erasure with quasi-static control strategies \citep{Berut2012,Jun2014,Berut2015,Gavrilov2016,Dago2021}.
The state lag accumulated in a finite-rate operation hinders the development
of finite-time erasure schemes. Our strategy offers an operable approach
to eliminate the nonequilibrium lag and realize the finite-time memory
erasure with high accuracy and low energy costs. Besides, the Landauer's
bound have been approached in an underdamped micromechanical oscillator
\citep{Dago2021,Dago2022}. The fast equilibrium recovery strategy
have also been achieved with a levitated particle in the underdamped
regime \citep{Raynal2023}. The driving force in our shortcut scheme
only depends on the particle's position. Therefore, it is promising
to realize our finite-time memory erasure strategy with current experimental
platforms.

\emph{Acknowledgment.}--This work is supported by the National Natural
Science Foundation of China (NSFC) (Grants No. 12088101, No. 11534002,
No. 11875049, No. U1930402, No. U1930403 and No. 12047549) and the
National Basic Research Program of China (Grant No. 2016YFA0301201).

\bibliographystyle{apsrev4-1}
\bibliography{ref}

\end{document}


\title{Supplementary Material: A Shortcut to Finite-time Memory Erasure}
\author{Geng Li}
\affiliation{Graduate School of China Academy of Engineering Physics, Beijing 100193,
China}
\affiliation{School of Systems Science, Beijing Normal University, Beijing 100875,
China}
\author{Hui Dong}
\email{hdong@gscaep.ac.cn}

\affiliation{Graduate School of China Academy of Engineering Physics, Beijing 100193,
China}

\maketitle
The supplementary materials are devoted to provide detailed derivations
in the main context.

\tableofcontents{}

\section{The auxiliary control in shortcuts to memory erasure\label{SSec-one}}

In this section, we will accomplish the finite-time memory erasure
with the shortcut scheme. The momentum-dependent terms in the auxiliary
control are removed by employing a variational method and a gauge
transformation.

\subsection{Approximate shortcut scheme\label{SSec-one-A}}

We consider a one-bit memory system described by the Hamiltonian $H_{o}(x,p,\vec{\lambda})\equiv p^{2}/(2m)+U_{o}(x,\vec{\lambda})$
with the modulated double-well potential $U_{o}(x,\vec{\lambda})=kx^{4}-A\lambda_{1}x^{2}-B\lambda_{2}x$.
In the shortcut scheme, an auxiliary Hamiltonian $H_{a}$ is added
to steer the system evolving along the instantaneous equilibrium state
$P_{\mathrm{eq}}=\exp[\beta(F-H_{o})]$ during the finite-time erasure
process with prescribed boundary conditions $H_{a}(0)=H_{a}(\tau)=0$.
The probability distribution $P(x,p,t)$ of the microstate evolves
according to the Kramers equation:

\begin{equation}
\frac{\partial P}{\partial t}=-\frac{\partial}{\partial x}(\frac{\partial H}{\partial p}P)+\frac{\partial}{\partial p}(\frac{\partial H}{\partial x}P+\gamma\frac{\partial H}{\partial p}P)+\frac{\gamma}{\beta}\frac{\partial^{2}P}{\partial p^{2}}],\label{seq:kramerseq}
\end{equation}
where $H\equiv H_{o}+H_{a}$ represents the total Hamiltonian. The
detailed derivation of the Kramers equation in the shortcut scheme
is presented in Reference$\ $\cite{Li2022}. With the demand of instantaneous
equilibrium paths, we derive the evolution equation for the auxiliary
control:

\begin{equation}
\frac{\gamma}{\beta}\frac{\partial^{2}H_{a}}{\partial p^{2}}-\frac{\gamma p}{m}\frac{\partial H_{a}}{\partial p}+\frac{\partial H_{a}}{\partial p}\frac{\partial H_{o}}{\partial x}-\frac{p}{m}\frac{\partial H_{a}}{\partial x}=\frac{dF}{dt}-\frac{\partial H_{o}}{\partial t},\label{seq:addilHamieq}
\end{equation}
where the auxiliary Hamiltonian $H_{a}(x,p,t)$ is proved to take
the form $H_{a}=\dot{\vec{\lambda}}\cdot\vec{f}$. The boundary conditions
of the auxiliary Hamiltonian $H_{a}$ are satisfied with the request
of $\dot{\vec{\lambda}}(0)=\dot{\vec{\lambda}}(\tau)=0$.

With the modulated double-well potential $U_{o}(x,\vec{\lambda})$,
it seems unlikely to solve Eq.$\ $(\ref{seq:addilHamieq}) for the
auxiliary Hamiltonian $H_{a}$ analytically. We use a variational
method \cite{Li2021,Li2023} to obtain an approximate auxiliary control
$H_{a}^{*}=\dot{\vec{\lambda}}\cdot\vec{f}^{*}(x,p,\vec{\lambda})$,
where $\vec{f}^{*}$ denotes an approximation to $\vec{f}$. In the
variational shortcut scheme, a functional is used to evaluate the
approximation of the auxiliary control $H_{a}^{*}$. The variational
functional is defined from Eq.$\ $(\ref{seq:addilHamieq}) as

\begin{equation}
\mathcal{G}(H_{a}^{*})=\int dxdp(\frac{\gamma}{\beta}\frac{\partial^{2}H_{a}^{*}}{\partial p^{2}}-\frac{\gamma p}{m}\frac{\partial H_{a}^{*}}{\partial p}+\frac{\partial H_{o}}{\partial x}\frac{\partial H_{a}^{*}}{\partial p}-\frac{p}{m}\frac{\partial H_{a}^{*}}{\partial x}+\frac{\partial H_{o}}{\partial t}-\frac{\mathrm{d}F}{\mathrm{d}t})^{2}\mathrm{e}^{-\beta H_{o}}.\label{seq:noncons}
\end{equation}
The best possible form of the approximate auxiliary control is achieved
from the variational equation $\delta\mathcal{G}(H_{\mathrm{a}}^{*})/\delta H_{\mathrm{a}}^{*}=0$.

Another difficulty of the underdamped shortcut scheme is the momentum-dependence
in the auxiliary control. The momentum-dependent driving force is
hard to be realized in the experimental setup due to the request to
constantly monitor the particle's speed \cite{GueryOdelin2019,GueryOdelin2023}.
With the help of the variational method and a gauge transformation,
we remove the momentum-dependent terms in the auxiliary control. We
firstly remove the high-order momentum-dependent terms by using the
variational method. We assume that the variational auxiliary Hamiltonian
takes the form as

\begin{equation}
H_{\mathrm{a}}^{*}=\dot{\lambda}_{1}(a_{4}xp+a_{3}p+a_{2}x^{2}+a_{1}x)+\dot{\lambda}_{2}(b_{4}xp+b_{3}p+b_{2}x^{2}+b_{1}x),\label{seq:apahamton}
\end{equation}
where $a_{n}\equiv a_{n}(\vec{\lambda})$ and $b_{n}\equiv b_{n}(\vec{\lambda})$
with $n=1,2,3,4$ are functions to be determined in the variational
procedure. With the specified form of the auxiliary control in Eq.$\ $(\ref{seq:apahamton}),
the variational operation over $H_{\mathrm{a}}^{*}$ is converted
to the partial-derivative operation over parameters $a_{n}\equiv a_{n}(\vec{\lambda})$
and $b_{n}\equiv b_{n}(\vec{\lambda})$ with $n=1,2,3,4$.

\subsection{An equivalent process with momentum-independent auxiliary control\label{SSec-one-B}}

Note that there are still linear momentum-dependent terms in the variational
auxiliary Hamiltonian (\ref{seq:apahamton}). In the variational shortcut
scheme employing the Hamiltonian $H=H_{o}+H_{a}^{*}$, the particle's
evolution is governed by the Langevin equation
\begin{eqnarray}
\dot{x} & = & \frac{p}{m}+\dot{\lambda}_{1}(a_{4}x+a_{3})+\dot{\lambda}_{2}(b_{4}x+b_{3}),\nonumber \\
\dot{p} & = & -4kx^{3}+2A\lambda_{1}x+B\lambda_{2}-\dot{\lambda}_{1}(a_{4}p+2a_{2}x+a_{1})-\dot{\lambda}_{2}(b_{4}p+2b_{2}x+b_{1})-\gamma\dot{x}+\xi(t),\label{seq:orlangea}
\end{eqnarray}
where $\xi(t)$ represents the Gaussian white noise. The measurement
difficulties of the momentum-dependent forces in Eq.$\ $(\ref{seq:orlangea})
impede practical implementations of the shortcut scheme experimentally
\cite{GueryOdelin2019,GueryOdelin2023}.

We introduce a gauge transformation $X=x$ and $P=p+m\dot{\lambda}_{1}(a_{4}x+a_{3})+m\dot{\lambda}_{2}(b_{4}x+b_{3})$
to obtain an equivalent process with the viarables $X$ and $P$ under
the evolution as follows
\begin{eqnarray}
\dot{X} & = & \frac{P}{m},\nonumber \\
\dot{P} & = & -\frac{\partial U_{o}}{\partial X}-\frac{\partial U_{a}}{\partial X}-\gamma\dot{X}+\xi(t).\label{seq:equiproceq}
\end{eqnarray}
Here the auxiliary potential for these new viarables is $U_{a}(X,t)=C_{2}(t)X^{2}+C_{1}(t)X$
with the parameters
\begin{eqnarray}
C_{2}(t) & = & \dot{\lambda}_{1}a_{2}+\dot{\lambda}_{2}b_{2}-\frac{m}{2}(\ddot{\lambda}_{1}a_{4}+\ddot{\lambda}_{2}b_{4}+\dot{\lambda}_{1}^{2}\frac{\partial a_{4}}{\partial\lambda_{1}}+\dot{\lambda}_{1}^{2}a_{4}^{2}+\dot{\lambda}_{2}^{2}\frac{\partial b_{4}}{\partial\lambda_{2}}+\dot{\lambda}_{2}^{2}b_{4}^{2}+\dot{\lambda}_{1}\dot{\lambda}_{2}\frac{\partial a_{4}}{\partial\lambda_{2}}+\dot{\lambda}_{1}\dot{\lambda}_{2}\frac{\partial b_{4}}{\partial\lambda_{1}}+2\dot{\lambda}_{1}\dot{\lambda}_{2}a_{4}b_{4}),\nonumber \\
C_{1}(t) & = & \dot{\lambda}_{1}a_{1}+\dot{\lambda}_{2}b_{1}-m(\ddot{\lambda}_{1}a_{3}+\ddot{\lambda}_{2}b_{3}+\dot{\lambda}_{1}^{2}\frac{\partial a_{3}}{\partial\lambda_{1}}+\dot{\lambda}_{1}^{2}a_{3}a_{4}+\dot{\lambda}_{2}^{2}\frac{\partial b_{3}}{\partial\lambda_{2}}+\dot{\lambda}_{2}^{2}b_{3}b_{4}+\dot{\lambda}_{1}\dot{\lambda}_{2}\frac{\partial a_{3}}{\partial\lambda_{2}}+\dot{\lambda}_{1}\dot{\lambda}_{2}\frac{\partial b_{3}}{\partial\lambda_{1}}\nonumber \\
 &  & +\dot{\lambda}_{1}\dot{\lambda}_{2}b_{3}a_{4}+\dot{\lambda}_{1}\dot{\lambda}_{2}a_{3}b_{4}).\label{seq:c1c2parame}
\end{eqnarray}
With the operation of Jacobi's transformation, we find that the system's
distribution in this equivalent process follows a fixed pattern as
\begin{equation}
P_{f}(X,P,t)=\exp\left\{ \beta[F-\frac{1}{2m}(P-m\partial H_{a}^{*}/\partial P)^{2}-U_{o}]\right\} .\label{seq:fixedpattern}
\end{equation}
Compared with the instantaneous equilibrium distribution $P_{\mathrm{eq}}=\exp[\beta(F-H_{o})]$,
the extra term $m\partial H_{a}^{*}/\partial P$ in the fixed distribution
$P_{f}$ vanishes with the consideration of the boundary conditions
$\dot{\vec{\lambda}}(0)=\dot{\vec{\lambda}}(\tau)=0$. Therefore,
the system's distribution $P_{f}$ returns to the instantaneous equilibrium
distribution $P_{\mathrm{eq}}$ at the beginning $t=0$ and end $t=\tau$
of the equivalent process with the Hamiltonian $H=H_{o}+U_{a}$.

\section{Dimensionless form of the erasure process\label{SSec-two}}

In this section, we numerically solve the function $f^{*}$ for the
auxiliary control. For later simulations, we introduce the characteristic
length $l_{\mathrm{c}}\equiv(k_{\mathrm{B}}T/k)$, the characteristic
times $\tau_{1}=m/\gamma$ and $\tau_{2}=\gamma/(kl_{c}^{2})$ to
define the dimensionless coordinate $\tilde{x}\equiv x/l_{\mathrm{c}}$,
momentum $\tilde{p}\equiv p\tau_{2}/(ml_{\mathrm{c}})$, time $s\equiv t/\tau_{2}$,
and the parameters $\tilde{A}\equiv A/(kl_{\mathrm{c}}^{2})$ and
$\tilde{B}\equiv B/(kl_{\mathrm{c}}^{3})$. The dimensionless evolution
equation for the auxiliary control follows as

\begin{equation}
\frac{1}{\alpha^{2}}\frac{\partial^{2}\tilde{H}_{a}}{\partial\tilde{p}^{2}}-\frac{1}{\alpha}\tilde{p}\frac{\partial\tilde{H}_{a}}{\partial\tilde{p}}+\frac{1}{\alpha}\frac{\partial\tilde{H}_{a}}{\partial\tilde{p}}\frac{\partial\tilde{H}_{o}}{\partial\tilde{x}}-\tilde{p}\frac{\partial\tilde{H}_{a}}{\partial\tilde{x}}=\frac{d\tilde{F}}{ds}-\frac{\partial\tilde{H}_{o}}{\partial s},\label{seq:dimleailHi}
\end{equation}
with the dimensionless parameter $\alpha\equiv\tau_{1}/\tau_{2}$
and the dimensionless Hamiltonian $\tilde{H}_{o}\equiv H_{o}/(k_{B}T)$
and $\tilde{H}_{a}\equiv H_{a}/(k_{B}T)$. The dimensionless variational
functional takes the form

\begin{equation}
\mathcal{\tilde{G}}(\tilde{H}_{a}^{*})=\int d\tilde{x}d\tilde{p}(\frac{1}{\alpha^{2}}\frac{\partial^{2}\tilde{H}_{a}}{\partial\tilde{p}^{2}}-\frac{1}{\alpha}\tilde{p}\frac{\partial\tilde{H}_{a}}{\partial\tilde{p}}+\frac{1}{\alpha}\frac{\partial\tilde{H}_{a}}{\partial\tilde{p}}\frac{\partial\tilde{H}_{o}}{\partial\tilde{x}}-\tilde{p}\frac{\partial\tilde{H}_{a}}{\partial\tilde{x}}+\frac{\partial\tilde{H}_{o}}{\partial s}-\frac{\partial\tilde{H}_{o}}{\partial s})^{2}\mathrm{e}^{-\tilde{H}_{o}},\label{seq:dinoncons}
\end{equation}
with the dimensionless variational auxiliary Hamiltonian

\begin{equation}
\tilde{H}_{\mathrm{a}}^{*}=\lambda_{1}^{'}(\tilde{a}_{4}\tilde{x}\tilde{p}+\tilde{a}_{3}\tilde{p}+\tilde{a}_{2}\tilde{x}^{2}+\tilde{a}_{1}\tilde{x})+\lambda_{2}^{'}(\tilde{b}_{4}\tilde{x}\tilde{p}+\tilde{b}_{3}\tilde{p}+\tilde{b}_{2}\tilde{x}^{2}+\tilde{b}_{1}\tilde{x}).\label{seq:diapahamton}
\end{equation}
Here $\lambda_{1}^{'}\equiv d\lambda_{1}/ds$, $\lambda_{2}^{'}\equiv d\lambda_{2}/ds$
and the parameters $\tilde{a}_{n}$ and $\tilde{b}_{n}$ are the dimensionless
version of $a_{n}$ and $b_{n}$ with $n=1,2,3,4$.

With the variational procedure of $\mathcal{\tilde{G}}(\tilde{H}_{a}^{*})$
over the parameters $\tilde{a}_{n}$ and $\tilde{b}_{n}$, we obtain
the set of equations as

\begin{eqnarray}
[48\langle\tilde{x}^{4}\rangle+(\frac{1}{\alpha}-16\lambda_{1})\langle\tilde{x}^{2}\rangle+3]\tilde{a}_{4}+[48\langle\tilde{x}^{3}\rangle+(\frac{1}{\alpha}-16\lambda_{1})\langle\tilde{x}\rangle]\tilde{a}_{3}+2\langle\tilde{x}^{2}\rangle\tilde{a}_{2}+\langle\tilde{x}\rangle\tilde{a}_{1} & = & 16\alpha\langle\tilde{x}^{2}\rangle,\nonumber \\{}
[48\langle\tilde{x}^{3}\rangle+(\frac{1}{\alpha}-16\lambda_{1})\langle\tilde{x}\rangle]\tilde{a}_{4}+[48\langle\tilde{x}^{2}\rangle+(\frac{1}{\alpha}-16\lambda_{1})]\tilde{a}_{3}+2\langle\tilde{x}\rangle\tilde{a}_{2}+\tilde{a}_{1} & = & 16\alpha\langle\tilde{x}\rangle,\nonumber \\
\frac{\langle\tilde{x}^{2}\rangle}{\alpha}\tilde{a}_{4}+\frac{\langle\tilde{x}\rangle}{\alpha}\tilde{a}_{3}+2\langle\tilde{x}^{2}\rangle\tilde{a}_{2}+\langle\tilde{x}\rangle\tilde{a}_{1} & = & 0,\nonumber \\
\frac{\langle\tilde{x}\rangle}{\alpha}\tilde{a}_{4}+\frac{1}{\alpha}\tilde{a}_{3}+2\langle\tilde{x}\rangle\tilde{a}_{2}+\tilde{a}_{1} & = & 0,\nonumber \\{}
[48\langle\tilde{x}^{4}\rangle+(\frac{1}{\alpha}-16\lambda_{1})\langle\tilde{x}^{2}\rangle+3]\tilde{b}_{4}+[48\langle\tilde{x}^{3}\rangle+(\frac{1}{\alpha}-16\lambda_{1})\langle\tilde{x}\rangle]\tilde{b}_{3}+2\langle\tilde{x}^{2}\rangle\tilde{b}_{2}+\langle\tilde{x}\rangle\tilde{b}_{1} & = & 16\alpha\langle\tilde{x}\rangle,\nonumber \\{}
[48\langle\tilde{x}^{3}\rangle+(\frac{1}{\alpha}-16\lambda_{1})\langle\tilde{x}\rangle]\tilde{b}_{4}+[48\langle\tilde{x}^{2}\rangle+(\frac{1}{\alpha}-16\lambda_{1})]\tilde{b}_{3}+2\langle\tilde{x}\rangle\tilde{b}_{2}+\tilde{b}_{1} & = & 16\alpha,\nonumber \\
\frac{\langle\tilde{x}^{2}\rangle}{\alpha}\tilde{b}_{4}+\frac{\langle\tilde{x}\rangle}{\alpha}\tilde{b}_{3}+2\langle\tilde{x}^{2}\rangle\tilde{b}_{2}+\langle\tilde{x}\rangle\tilde{b}_{1} & = & 0,\nonumber \\
\frac{\langle\tilde{x}\rangle}{\alpha}\tilde{b}_{4}+\frac{1}{\alpha}\tilde{b}_{3}+2\langle\tilde{x}\rangle\tilde{b}_{2}+\tilde{b}_{1} & = & 0,\label{seq:setparaequ}
\end{eqnarray}
where $\langle\tilde{x}^{n}\rangle\equiv\int_{-\infty}^{+\infty}\tilde{x}^{n}\exp(-\tilde{U}_{o})d\tilde{x}/\int_{-\infty}^{+\infty}\exp(-\tilde{U}_{o})d\tilde{x}$
with $n=1,2,3,4$. We first numerically obtain the values of $\langle\tilde{x}^{n}\rangle$
for different values of control parameters $\lambda_{1}$ and $\lambda_{2}$.
According to the boundary conditions of $\lambda_{1}$ and $\lambda_{2}$,
we set the range of control parameters to be $\lambda_{1}\in[-5,5]$
and $\lambda_{2}\in[-5,5]$. Second, we numerically solve the set
of Eqs.$\ $(\ref{seq:setparaequ}) for different values of $\langle\tilde{x}^{n}\rangle$
marked by different set of control parameters $\lambda_{1}$ and $\lambda_{2}$.
In this way, we obtain the corresponding solutions of $\tilde{a}_{n}$
and $\tilde{b}_{n}$ for different set of control parameters $\lambda_{1}$
and $\lambda_{2}$. Finally, we fit the polynomial relation between
the coefficients $\tilde{a}_{n}$, $\tilde{b}_{n}$ and the parameters
$\lambda_{1}$, $\lambda_{2}$.

The dimensionless auxiliary potential takes the form as
\begin{eqnarray}
\tilde{U}_{a}(\tilde{X},s) & = & \tilde{C}_{2}(s)\tilde{X}^{2}+\tilde{C}_{1}(s)\tilde{X},\label{seq:dimlessauxpot}
\end{eqnarray}
where the dimensionless parameters follow as
\begin{eqnarray}
\tilde{C}_{2}(s) & = & \lambda_{1}^{'}\tilde{a}_{2}+\lambda_{2}^{'}\tilde{b}_{2}-\frac{1}{2}(\lambda_{1}^{''}\tilde{a}_{4}+\lambda_{2}^{''}\tilde{b}_{4}+\lambda_{1}^{'2}\frac{\partial\tilde{a}_{4}}{\partial\lambda_{1}}+\frac{1}{\alpha}\lambda_{1}^{'2}\tilde{a}_{4}^{2}+\lambda_{2}^{'2}\frac{\partial\tilde{b}_{4}}{\partial\lambda_{2}}+\frac{1}{\alpha}\lambda_{2}^{'2}\tilde{b}_{4}^{2}+\lambda_{1}^{'}\lambda_{2}^{'}\frac{\partial\tilde{a}_{4}}{\partial\lambda_{2}}+\lambda_{1}^{'}\lambda_{2}^{'}\frac{\partial\tilde{b}_{4}}{\partial\lambda_{1}}+\frac{2}{\alpha}\lambda_{1}^{'}\lambda_{2}^{'}\tilde{a}_{4}\tilde{b}_{4}),\nonumber \\
\tilde{C}_{1}(s) & = & \lambda_{1}^{'}\tilde{a}_{1}+\lambda_{2}^{'}\tilde{b}_{1}-\lambda_{1}^{''}\tilde{a}_{3}-\lambda_{2}^{''}\tilde{b}_{3}-\lambda_{1}^{'2}\frac{\partial\tilde{a}_{3}}{\partial\lambda_{1}}-\frac{1}{\alpha}\lambda_{1}^{'2}\tilde{a}_{3}\tilde{a}_{4}-\lambda_{2}^{'2}\frac{\partial\tilde{b}_{3}}{\partial\lambda_{2}}-\frac{1}{\alpha}\lambda_{2}^{'2}\tilde{b}_{3}\tilde{b}_{4}-\lambda_{1}^{'}\lambda_{2}^{'}\frac{\partial\tilde{a}_{3}}{\partial\lambda_{2}}-\lambda_{1}^{'}\lambda_{2}^{'}\frac{\partial\tilde{b}_{3}}{\partial\lambda_{1}}\nonumber \\
 &  & -\frac{1}{\alpha}\lambda_{1}^{'}\lambda_{2}^{'}\tilde{b}_{3}\tilde{a}_{4}-\frac{1}{\alpha}\lambda_{1}^{'}\lambda_{2}^{'}\tilde{a}_{3}\tilde{b}_{4}).\label{seq:dimlessparame}
\end{eqnarray}

\section{Geodesic path for shortcuts to memory erasure\label{SSec-three}}

In this section, we use geometric approach to find the geodesic path
with minimal energy costs for the erasure process controlled by the
shortcut scheme. With the variational auxiliary Hamiltonian in Eq.$\ $(\ref{seq:diapahamton}),
we obtain the dimensionless metric for the parametric space as
\begin{equation}
\tilde{g}=\left(\begin{array}{cc}
\tilde{a}_{4}^{2}\langle\tilde{x}^{2}\rangle+2\tilde{a}_{3}\tilde{a}_{4}\langle\tilde{x}\rangle+\tilde{a}_{3}^{2} & \tilde{a}_{4}\tilde{b}_{4}\langle\tilde{x}^{2}\rangle+(\tilde{a}_{3}\tilde{b}_{4}+\tilde{b}_{3}\tilde{a}_{4})\langle\tilde{x}\rangle+\tilde{a}_{3}\tilde{b}_{3}\\
\tilde{a}_{4}\tilde{b}_{4}\langle\tilde{x}^{2}\rangle+(\tilde{a}_{3}\tilde{b}_{4}+\tilde{b}_{3}\tilde{a}_{4})\langle\tilde{x}\rangle+\tilde{a}_{3}\tilde{b}_{3} & \tilde{b}_{4}^{2}\langle\tilde{x}^{2}\rangle+2\tilde{b}_{3}\tilde{b}_{4}\langle\tilde{x}\rangle+\tilde{b}_{3}^{2}
\end{array}\right).\label{seq:geometric}
\end{equation}
The polynomial relation between the metric $\tilde{g}_{\mu\nu}$ and
the parameters $\lambda_{1}$, $\lambda_{2}$ is fitted by substituting
the values of $\tilde{a}_{n}$, $\tilde{b}_{n}$, and $\langle\tilde{x}^{n}\rangle$
into Eq.$\ $(\ref{seq:geometric}).

According to the geometric approach presented in the main text, the
task of finding optimal erasure protocol with minimal energy costs
is equivalent to solving the geodesic path in the parametric space
with the metric $\tilde{g}_{\mu\nu}$. The geodesic path is obtained
by solving the geodesic equation
\begin{equation}
\ddot{\lambda}_{\mu}+\sum_{\nu\kappa}\Gamma_{\nu\kappa}^{\mu}\dot{\lambda}_{\nu}\dot{\lambda}_{\kappa}=0,\label{seq:geodesiceq}
\end{equation}
with the given boundary conditions $\vec{\lambda}(0)=(1,0)$ and $\vec{\lambda}(\tau)=(0,1)$.
Here the Christoffel symbol is defined as $\Gamma_{\nu\kappa}^{\mu}\equiv\frac{1}{2}\sum_{\iota}(\tilde{g}^{-1})_{\iota\mu}(\partial_{\lambda_{\kappa}}\tilde{g}_{\iota\nu}+\partial_{\lambda_{\nu}}\tilde{g}_{\iota\kappa}-\partial_{\lambda_{\iota}}\tilde{g}_{\nu\kappa})$.
We employ the shooting method \cite{Berger2007} to numerically solve
the geodesic equation. The two-point boundary values problem is treated
as an initial value problem

\begin{equation}
\ddot{\lambda}_{\mu}=y_{\mu}(t,\vec{\lambda},\dot{\vec{\lambda}})\equiv\frac{1}{2}\sum_{\nu\kappa\iota}(g^{-1})_{\iota\mu}(\frac{\partial g_{\nu\kappa}}{\partial\lambda_{\iota}}-\frac{\partial g_{\iota\nu}}{\partial\lambda_{\kappa}}-\frac{\partial g_{\iota\kappa}}{\partial\lambda_{\nu}})\dot{\lambda}_{\nu}\dot{\lambda}_{\kappa},\label{seq:vargeodequ}
\end{equation}
with the initial conditions $\vec{\lambda}(0)=(1,0)$ and $\dot{\vec{\lambda}}(0+)=\vec{d}$.
Here $\vec{d}$ is the initial rate that is constantly updated until
the solution of Eq.$\ $(\ref{seq:vargeodequ}) reaches the required
position $\vec{\lambda}(\tau)=(0,1)$. In the simulation, we use the
Eular algorithm to solve the geodesic equation$\ $(\ref{seq:vargeodequ}).
The Newton's method is taken to iterate the initial rate $\vec{d}$
and finally satisfy the boundary condition $\vec{\lambda}(\tau)=(0,1)$.

\section{The stochastic simulations\label{SSec-four}}

In this section, we introduce the algorithm to simulate the erasure
process and obtain the mean work. The motion of the memory system
is described by the Langevin equation in Eq.$\ $(\ref{seq:equiproceq}).
The dimensionless version of this equation follows as
\begin{eqnarray}
\tilde{X}' & = & \tilde{P},\nonumber \\
\tilde{P}' & = & -\frac{1}{\alpha}\frac{\partial\tilde{U}_{o}}{\partial\tilde{X}}-\frac{1}{\alpha}\frac{\partial\tilde{U}_{a}}{\partial\tilde{X}}-\frac{\tilde{P}}{\alpha}+\frac{\sqrt{2}}{\alpha}\zeta(s)\text{,}\label{seq:moverdamped}
\end{eqnarray}
where $\zeta(s)$ is a Gaussian white noise satisfying $\langle\zeta(s)\rangle=0$
and $\langle\zeta(s_{1})\zeta(s_{2})\rangle=\delta(s_{1}-s_{2})$.
The Langevin equation~(\ref{seq:moverdamped}) is solved by employing
the Euler algorithm as
\begin{eqnarray}
\tilde{X}(s+\delta s) & = & \tilde{X}(s)+\tilde{P}\delta s,\nonumber \\
\tilde{P}(s+\delta s) & = & \tilde{P}(s)-\frac{1}{\alpha}\frac{\partial\tilde{U}_{o}}{\partial\tilde{X}}\delta s-\frac{1}{\alpha}\frac{\partial\tilde{U}_{a}}{\partial\tilde{X}}\delta s-\frac{\tilde{P}}{\alpha}\delta s+\frac{\sqrt{2\delta s}}{\alpha}\theta(s),\label{seq:simulatlange}
\end{eqnarray}
where $\delta s$ is the time step and $\theta(s)$ is a random number
sampled from Gaussian distribution with zero mean and unit variance.
The work of the system's stochastic trajectory follows as
\begin{eqnarray}
\tilde{w}\equiv\frac{w}{k_{B}T} & = & \int_{0}^{1}(\frac{\partial\tilde{H}_{o}}{\partial s}+\frac{\partial\tilde{U}_{a}}{\partial s})ds\nonumber \\
 & \approx & \sum(\frac{\partial\tilde{H}_{o}}{\partial s}+\frac{\partial\tilde{U}_{a}}{\partial s})\delta s.\label{seq:-moverdampedtrw}
\end{eqnarray}
In the simulation, we have chosen the parameters $\vec{\lambda}(0)=(1,0)$,
$\vec{\lambda}(\tau)=(0,1)$, $k_{\mathrm{B}}T=1$, $\gamma=1$, and
$m=0.01$. The mean work is obtained as the ensemble average over
the trajectory work of $10^{5}$ stochastic trajectories.

\bibliographystyle{unsrt}
\bibliography{ref}